# CentOS Linux for the ATLAS MUCTPI Upgrade

R. Spiwoks, A. Armbruster, P. Czodrowski, N. Ellis, P. Farthouat, S. Haas, A. Kulinska, A. Marzin,
P. Papageorgiou, T. Pauly, S. Perrella, M. Saimpert, P. Vichoudis, T. Wengler

*Abstract*—A new Muon-to-Central-Trigger Processor Interface (MUCTPI) was built as part of the upgrade of the ATLAS Level-1 trigger system for the next Run of the Large Hadron Collider at CERN. The MUCTPI has 208 high-speed optical serial links for receiving muon candidates from the muon trigger detectors. Three high-end FPGAs are used for real-time processing of the muon candidates, for sending trigger information to other parts of the trigger system, and for sending summary information to the data acquisition and monitoring system. A System-on-Chip (SoC) is used for the control, configuration and monitoring of the hardware and the operation of the MUCTPI. The SoC consists of an FPGA part and a processor system. The FPGA part provides communication with the processing FPGAs, while the processor system runs software for communication with the run-control system of the ATLAS experiment. In this paper we will describe our experience with running CentOS Linux on the SoC. Cross-compilation together with the existing framework for building of the ATLAS trigger and data acquisition (TDAQ) software is being used in order to allow the deployment of the TDAQ software directly on the SoC.

## I. Introduction

ATLAS is a general-purpose experiment at the Large Hadron Collider (LHC) at CERN [1]. It observes proton-proton collisions at a center-of-mass energy of almost 14 TeV, with a bunch crossing (BC) every 25 ns (40 MHz) and up to 80 pile-up collisions expected for the next run (Run 3), starting in 2022. This results in more than $10^9$ interactions per second and requires the use of a trigger system in order to select the events most interesting for physics studies within constraints on the maximum rate that can be recorded.

The ATLAS trigger and data acquisition (TDAQ) system [2] consists of a two-level trigger system. The Level-1 trigger is based on custom electronics and firmware, and reduces the event rate to 100 kHz. The High-level trigger is based on commercial-off-the-shelf computers and network, and processing software, and reduces the event rate to around 1.5 kHz (peak) which corresponds to a data rate of around 1.5 GByte/s. The Level-1 trigger uses reduced-granularity information from the calorimeters and information from dedicated muon trigger detectors, the Resistive Plate Chambers (RPC) in the barrel region ($|\eta|<1.05$) and the Thin-Gap Chambers (TGC) in the endcap and forward region ($1.05<|\eta|<2.5$), where $\eta$ is pseudo-rapidity, see Fig. 1.

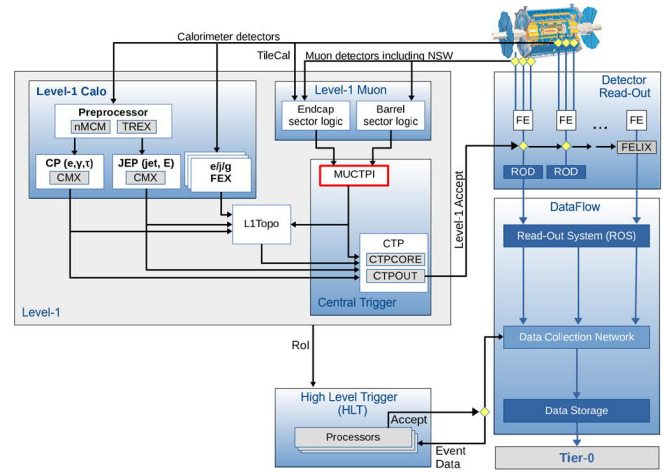

Fig. 1. Schematic of the ATLAS TDAQ systems [3].

The Muon-to-Central-Trigger-Processor Interface (MUCTPI) receives muon candidate information from 208 muon Sector-Logic modules, and combines the information. It avoids double counting of single muons, which are detected by more than one muon sector due to the geometrical overlap of the chambers and the trajectory of the muons in the magnetic field. This is called overlap handling. A typical overlap of a number of barrel and endcap sectors is shown in Fig. 2.

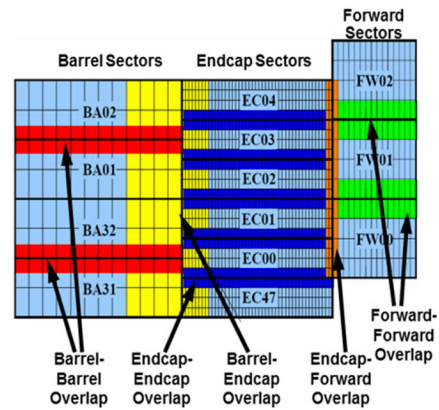

Fig. 2. Geometrical overlap of a typical region.

Up to sixteen highest transverse-momentum candidates after taking into account the overlap handling are sent to the topological trigger processor (L1Topo), while muon multiplicity information is sent to the Central Trigger Processor (CTP). The CTP combines all trigger information and takes the final Level-1 Accept or Reject decision.

## II. MUCTPI UPGRADE

For Run 3 of the LHC starting in 2022, higher rates of muons are expected. In order to enhance the trigger selectivity, more muon candidates per sector and more information per candidate will be sent, requiring higher bandwidth between the muon sector logics and the MUCTPI. Improved overlap handling allowing for possible overlap between octants that was previously not supported, will be provided. This requires more processing power. In addition, muon candidates with the full positional granularity will be sent to the topological trigger processor, and muon-only topological processing could possibly be carried out in the MUCTPI. Furthermore, the MUCTPI should be future proof, allowing for further improvements for Run 4 starting in 2027 [4].

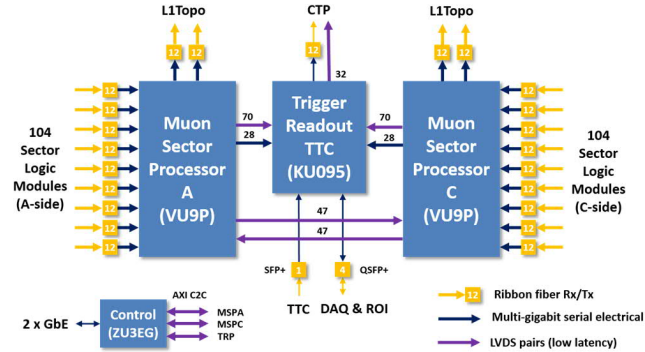

Fig. 3. Architecture of the MUCTPI Upgrade.

The previous VME-based system [5] with bulky electrical cables and 18 modules will be replaced by an ATCA-based system with optical cables and a single module using state-of-the-art FPGAs. Two Muon Sector Processors (MSPs) based on Xilinx Virtex Ultrascale+ FPGAs [6] receive the muon trigger information for the two halves of the complete system.

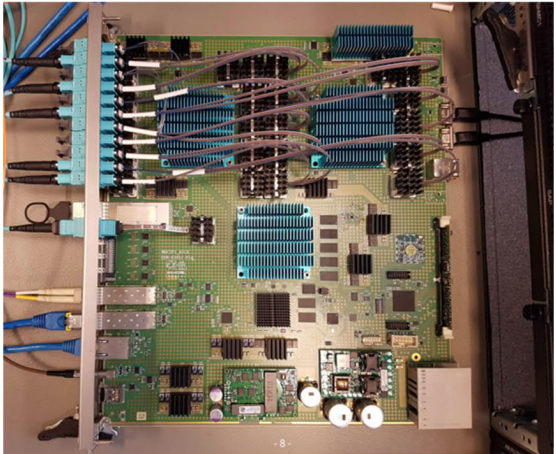

Fig. 4. Photograph of a prototype ATCA module for the MUCTPI Upgrade.

The MSPs implement the overlap handling and calculate multiplicities for each half of the ATLAS detector. This information is sent to the Trigger and Readout Processor (TRP) based on a Xilinx Kintex Ultrascale+ FPGA [6], which combines the partial results and sends overall multiplicities and other flags to the CTP. The two MSPs further send lists of muon candidates with full-precision information to the Topological Trigger Processor. The architecture of the upgraded MUCTPI is shown in Fig. 3, and a photograph of the current prototype is shown in Fig. 4.

## III. SYSTEM-ON-CHIP

In the previous VME-based system, a Single-Board Computer (SBC) in the VME crate controlled the MUCTPI. The SBC was based on x86_64 processors and was running Scientific Linux CERN (SLC) [7]. The user application software and its development was fully integrated into the ATLAS TDAQ system. In the new ATCA-based MUCTPI, a System-on-Chip (SoC) is used for control, configuration, and monitoring of the hardware and the operation of the MUCTPI, see Fig. 5.

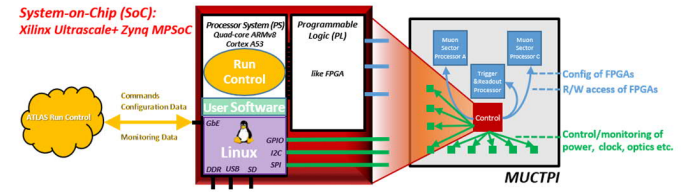

Fig. 5. Overview of the role of a System-on-Chip in the MUCTPI Upgrade.

The SoC consists of a processor system (PS), like a CPU, and a programmable logic (PL), like an FPGA. The PS of the SoC is based on multi-core ARM processors. It has memory and peripherals, like Gigabit Ethernet for communication with the ATLAS run-control system, and I2C, SPI, GPIO etc. for the control of the hardware. It runs software, which can be a "bare-metal" application or a full-blown operating system like Linux. The PL of the SoC has logic cells, memory block, IO links and Multi-Gigabit Transceivers (MGTs). It implements the interfaces to the other processing FPGAs, and can implement real-time logic or additional peripherals, e.g. 10 Gigabit Ethernet.

Since the MUCTPI uses Xilinx FPGAs, it was natural to use a Xilinx SoC, and a choice was made for the Xilinx Zynq Ultrascale+ MPSoC ZU3EG [6] which has a quad ARM Cortex A53 processor with a 64-bit architecture (aarch64), runs at 1.2 GHz and has 4 GByte of memory. While the Xilinx/Vivado and Xilinx/SDK tools [6] provide several of the files required for booting the SoC, like the bit stream file for the PL, the first-stage boot loader for initializing the hardware and for loading the boot loader, and the device tree file for use with the Linux kernel, there is a choice for the operating system. Xilinx PetaLinux [8], or the Yocto/OpenEmbedded framework [9] together with the Xilinx meta-layers could be used to build the Linux kernel, the Linux root file system, and the U-Boot for loading the kernel, device tree and root file system. However, this is not a CERN-certified operating system and it would not be allowed to run it on the ATLAS Technical Control Network (ATCN) of the experiment for security reasons. The SoC, and the MUCTPI, would therefore need to be isolated behind a gateway PC. In addition, providing this operating system would require continuous effort for updates, and the ATLAS system administrators or CERN-IT would not provide support for it. A different solution for the operating system therefore had to be found.

## IV. USE OF CENTOS LINUX

The operating system is key to the operation of the SoC and its integration into the ATCN and the ATLAS TDAQ software. We chose to use CentOS [10] because it is widely accepted at CERN, it is available for the aarch64 architecture, and is continuously updated by the CentOS community.

The CentOS/aarch64 operating system is installed on a host PC using the DNF cross installer [11]. This makes use of the QEMU emulator [12] and the Linux kernel support for miscellaneous binary formats [13]. The full recipe can be found at the web pages of the system-on-chip@cern.ch interest group [14], which held a workshop on SoC in 2019 [15] and which continues with a series of meetings [16]. The organizing committee is following up on issues of common interest, like the possible support of CentOS/aarch64 by CERN-IT in the same way as for the x86_64 systems. This is not yet agreed upon for Run 3, and will only become available for Run 4. In the meantime, the MUCTPI will still use CentOS and run isolated behind a gateway PC.

The cross-installed CentOS/aarch64 root file system is used together with a Linux kernel using the drivers and settings required by Xilinx, and the U-Boot loader built with the Yocto/OpenEmbedded framework - it could equally be built using Xilinx PetaLinux. The root file system is mounted using NFS for simplicity, but could also be based on an SD card. An overview of that setup is shown in Fig. 6.

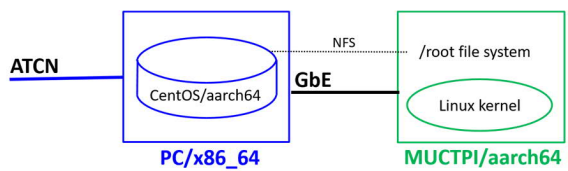

Fig. 6. Software architecture of the MUCPTI/SoC using a cross-installed CentOS/aarch64 root file system.

Booting the SoC as described above provides a full operating system with ssh for login, bash, python, tcl, expect, etc. for scripting, as well as NFS and NTP for mounting other file systems. Also available are Linux tools for networking like ip and ifconfig or a DHCP client, as well as the iptables tool for network security. Many other packages can be added to the installation easily. The result is that the SoC behaves exactly like any other CERN Linux system.

In order to add user application software accessing the MUCTPI's hardware features, and in order to integrate into the ATLAS TDAQ run-control system, cross-compilation is used. The gcc/aarch64 compiler from the CentOS repository is not necessarily up to date, but developer.arm.com or building from the source from gcc.gnu.org can be used to get an up to date cross-compiler.

Using CentOS as the system root, the software development flow from the previous MUCTPI generation can be re-used:
1. All registers and their bit fields, as well as memories and FIFOs are described in an XML file.
2. We developed a dedicated software tool, which takes the XML file and generates VHDL code for the firmware, as well as C++ code for the software, effectively providing firmware/software co-development.
3. The registers and memories are mapped into the PS part of the SoC using AXI, and can be accessed using /dev/uio files.
4. C++ code accesses the /dev/uio files for single read/write functions. We have also developed a kernel module for DMA.

The cross-compilation works within the framework of CMake [17], which is the build tool chosen by the ATLAS TDAQ and the MUCTPI teams. The ATLAS TDAQ software can also be cross-compiled in the same way. This allows us to build a run-control application, which can run on the SoC of the MUCTPI. Note that a gateway application running on the host PC is still required until the MUCTPI will be allowed directly on the ATCN. An overview of the software architecture is shown in Fig. 7.

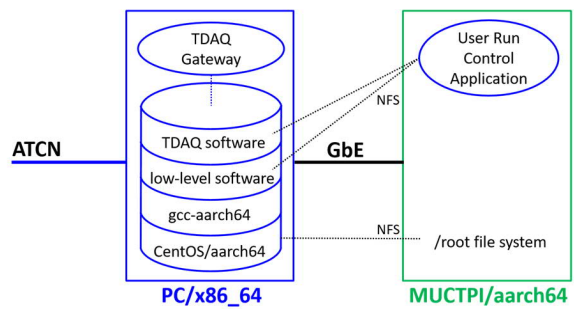

Fig. 7. Software architecture of the MUCPTI/SoC using cross-compiled user application software and a TDAQ Run-Control Application.

In addition, the Worldwide LHC Computing Grid (WLCG) [18] provides software builds for the aarch64 architecture, which allows us to use e.g. ROOT [19] as a library for histograms, graphs and other objects. In this way, the run-control application can produce histograms to be sent to the ATLAS TDAQ run-control system. Furthermore, we have developed a single setup script to set up the host PC for development, i.e. cross-compilation, or the SoC of the MUCTPI for deployment. In this way, the developer or user does not notice the "cross" environment and the flow of software development works exactly as before when using the VME-based SBC.

As a result, the MUCTPI C++ low-level software with its menu and test programs runs directly on the MUCTPI's SoC. In addition, Python wrappers for that software allow the firmware developers to write test scripts for interactive debugging. In addition, the MUCTPI C++ low-level software is used by an ATLAS TDAQ run-control application for control of the MUCTPI from the ATLAS TDAQ run-control system.

## V. Conclusion

A new MUCTPI was built in the ATCA form factor and using a System-on-Chip (SoC). The SoC is used for configuration, control and monitoring of the hardware and the operation of the MUCTPI. The SoC runs the CentOS Linux operating system. This is used together with cross-compilation for the SoC. User application software can be developed with a development flow which is exactly the same as before, when using a VME-based system and a Single-Board Computer.

It has also to be noted that there are many projects for the High-Luminosity LHC (HL-LHC) starting in 2027, in particular in the ATLAS and CMS experiments, which are planning to use SoCs. There are projections of several thousands of SoCs to be installed in the experiments. Using standard Linux operating systems will provide them with the network security and the system administration support necessary. The SoC interest group is a forum for looking into common solutions for such systems.

Acknowledgments

We would like to thank M. Wittgen, SLAC, USA, for the original idea of cross installing the CentOS root file system, and S. Kolos, University of California Irvine, USA, for providing the ATLAS TDAQ gateway application.